\documentclass[10pt,english,aps,amssymb,prl,twocolumn, showpacs]{revtex4-1}
\pdfoutput=1
\usepackage[utf8]{inputenc}
\usepackage{amsmath}
\usepackage{graphicx}
\usepackage{amssymb}
\usepackage{esint}
\usepackage{verbatim}

\makeatletter
\@ifundefined{textcolor}{}
{
 \definecolor{WHITE}{gray}{1}
 \definecolor{RED}{rgb}{1,0,0}
 \definecolor{GREEN}{rgb}{0,1,0}
 \definecolor{BLUE}{rgb}{0,0,1}
 \definecolor{CYAN}{cmyk}{1,0,0,0}
 \definecolor{MAGENTA}{cmyk}{0,1,0,0}
 \definecolor{YELLOW}{cmyk}{0,0,1,0}
 }

\newcommand{\bra}[1]{\langle #1|}
\newcommand{\ket}[1]{|#1\rangle}

\renewcommand{\phi}{\varphi}
\renewcommand{\epsilon}{\varepsilon}

\renewcommand{\vec}[1]{{\bf #1}}

\DeclareMathOperator{\IM}{Im}
\DeclareMathOperator{\RE}{Re}
\DeclareMathOperator{\TR}{Tr}

\DeclareMathOperator{\sgn}{sgn}

\newcommand{\bs}{\boldsymbol}
\newcommand{\mc}{\mathcal}

\usepackage{babel}
\begin{document}
\title {Superlattice platform for chiral superconductivity with tuneable and high Chern numbers}
\author{Kim Pöyhönen}
\author{Teemu Ojanen}
\affiliation{Department of Applied Physics (LTL), Aalto University, P.~O.~Box 15100,
FI-00076 AALTO, Finland }
\date{\today}
%
%
%
%
%
%
\begin{abstract}
Finding concrete realizations for  topologically nontrivial chiral superconductivity has been a long-standing goal in quantum matter research. Here we propose a route to a systematic realization of chiral superconductivity with nonzero Chern numbers. This goal can be achieved in a nanomagnet lattice deposited on top of a spin-orbit coupled two-dimensional electron gas (2DEG) with proximity $s$-wave superconductivity. The proposed structure can be regarded as a universal platform for chiral superconductivity supporting a large variety of topological phases. The topological state of the system can be electrically controlled by, for example, tuning the density of the 2DEG.     
\end{abstract}
\maketitle
\bigskip{}

\textit{Introduction ---}
The Bardeen-Cooper-Schrieffer (BCS) theory explains superconductivity in terms of paired electrons, Cooper pairs, that condense in the same quantum state with a macroscopic population. One of the most striking subsequent predictions was the fact that a condensate could carry net angular momentum, giving rise to macroscopic chirality. However, candidates for chiral superconductors are rare \cite{kallin}. From the modern point of view, 2d chiral superconductors are naturally discussed in the context of topologically nontrivial states of matter that is classified by the Chern number invariant \cite{volovik, bernevig,schnyder}. In intrinsic chiral superconductors the Chern number is fixed to a certain value determined by the microscopic form of interparticle interactions. 

In this work we introduce a universal platform for 2d topological superconductivity that realizes a large collection of states with distinct Chern numbers. The central elements of the studied system are a nanomagnet lattice deposited on two-dimensional electron gas (2DEG) with significant spin-orbit coupling which is made superconducting through the proximity effect. Importantly, fabrication of the studied system is within the reach of current  technology. Furthermore, this system is tuneable through structural design as well as by gate operation, which allows switching between the different topological states after the structure is fixed. 

The proposed nanomagnet structure generalizes conceptually and operationally the ferromagnet-2DEG-superconductor sandwich structure that was proposed as a realization of the chiral state with Chern number one \cite{sau}. Instead of the simplest nontrivial state, our model exhibits a large number distinct phases with multiple chiral Majorana edge states. The  flexible tuneability also enables edge-mode engineering through fabrication of topological phase boundaries in the system.  Therefore the studied system could serve as an ideal testbed for the Majorana edge modes. Motivated by studies of topological superconductivity in magnetic chains \cite{np2,ruby,choy, np,brau1,klin,vazifeh, pientka2,poyh,heimes,bry}, 2d superconductors with large Chern numbers were previously discovered in superconducting surfaces decorated by magnetic atoms \cite{ront1,ront2,li1}. The long-range hybridization of subgap Yu-Shiba-Rusinov states \cite{yu,shiba,rusinov,balatsky} generally gives rise to rich, mosaic-like topological phase diagrams \cite{ront1,ront2}. However, as a crucial difference to atomic systems, the presently studied system allows a high-level of control in the fabrication, tuning and operating the system.  


In the present paper we solve the subgap spectrum of a circular magnet on a superconducting 2DEG system. The magnetic lattice problem is then formulated in terms of the subgap states of individual nanomagnets. Then we solve the spectrum of 1d and 2d magnetic lattices and investigate their topological properties. In the 2d case we discover a remarkably rich topological phase diagram, where the energy gaps protecting the states can be a significant fraction of the induced gap in the 2DEG. Our results indicate that the studied system offers an unprecedented opportunity to  systematically probe chiral superconductivity in experimentally feasible  systems.

\begin{figure}
\includegraphics[width=0.9\linewidth]{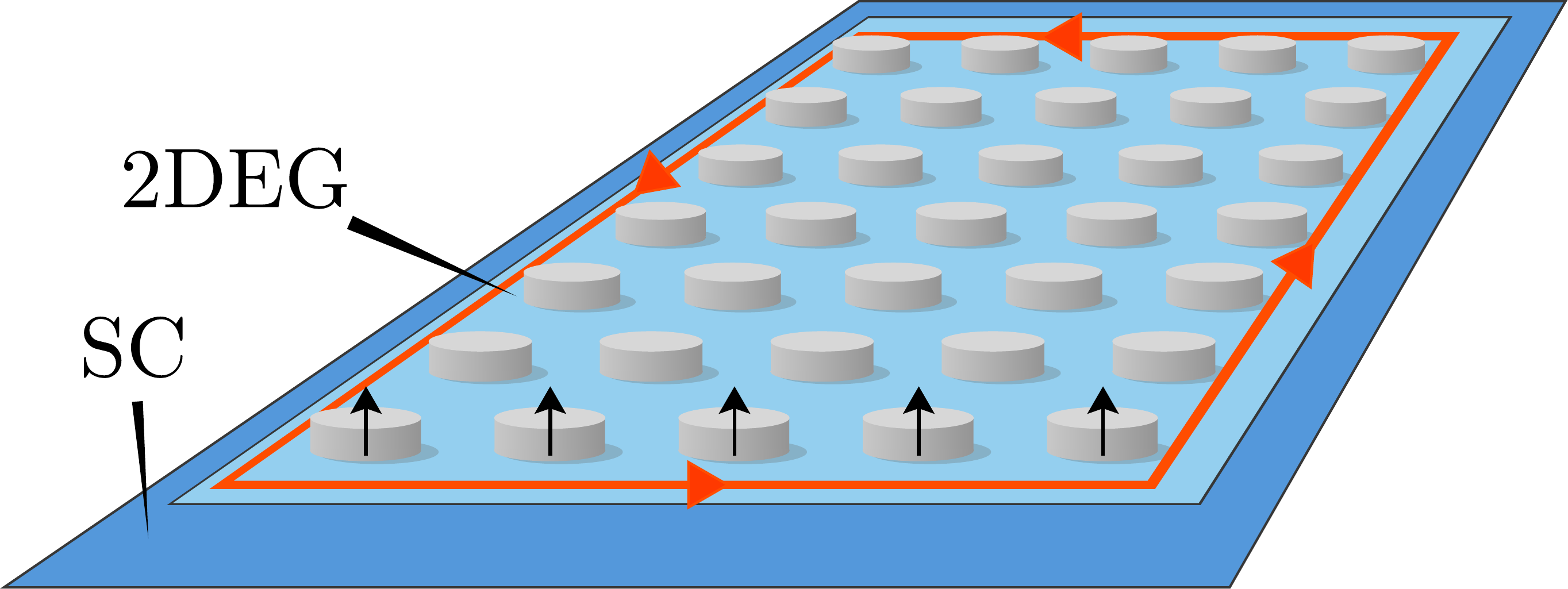}
\caption{An array of nanomagnets on a 2DEG substrate with spin-orbit coupling and proximity superconductivity realizes a highly tuneable platform of topological superconductivity with a large number of nontrivial states.}
\end{figure}

%
%
%
%
%
%
\textit{System --- }
We consider a 2DEG proximity coupled to an $s$-wave superconductor. In addition, a collection of nanomagnets have been placed on top of this substrate as shown in Fig.~1. The precise lattice geometry of the nanomagnet arrangement is not important for the derivation of the general description. In the standard Nambu basis
$\Psi = (\psi_\uparrow,\ \psi_\downarrow, \psi_\downarrow^\dagger,\ -\psi_\uparrow^\dagger)^T$, this system can be described by the $4\times 4$ Bogoliubov-de~Gennes Hamiltonian $H = H_0 + H_\text{imp}(\vec r)$, where
\begin{equation}
\begin{cases}
H_0 = \xi_k\tau_z + \alpha_R \vec k\times\bs\sigma\tau_z + \Delta \tau_x\\
H_\text{imp}(\vec r) = \sum_j V_j(\vec r - \vec r_j)
\end{cases}.\label{eq:bdghamiltonian}
\end{equation}
Here $\xi_k = \frac{k^2}{2m} - \mu$ is the kinetic energy, $\alpha_R$ the Rashba spin-orbit coupling, and $\Delta$ the induced superconducting order parameter. The matrices $\tau_i$ and $\sigma_i$ act in particle-hole and spin space, respectively.  The magnets are assumed be in a direct contact with the 2DEG, inducing a perpendicular Zeeman potential $V_j(\vec r)$, which takes the form 
\begin{equation}
V_j(\vec r) = M_j\sigma_z\theta(R_j -|\vec r|).\label{eq:magpotreal}
\end{equation} 
This corresponds to homogeneous circular magnets, each with a radius $R_i$ and magnetization energy scale $M_i$. In general, the magnets will also give rise to a local scalar potential. Since this effect only renormalizes the magnetic subgap states studied below, we will only consider the Zeeman part of the potential.  

In analogy to the Yu-Shiba-Rusinov states of magnetic atoms on a superconductor, a single nanomagnet gives rise to energy states penetrating in the gap \cite{balatsky}. Similarly to magnetic atom chains \cite{pientka2}, the topological properties of the nanomagnet lattice can be understood in terms of the subgap energy bands of hybridized bound states. We now wish to obtain an equation from which the energy bands and topological properties of the system can be discerned. From the equation $H\Psi = E\Psi$, by separating the magnetic potential on one side, we obtain the equation
\begin{equation}
\psi(\vec k) = \sum_jG_0(\vec k, E)\int \frac{d^2q}{(2\pi)^2} e^{-i \left(\vec k - \vec q\right) \cdot \vec r_j} V_j(\vec k - \vec q) \psi(\vec q),
\end{equation}
where $G_0(\vec k, E)=(E-H_0)^{-1}$. The dependence here on two momenta makes an exact solution challenging. To simplify the problem we follow the method of Refs.~\cite{kim,zhang} and proceed by assuming that the potential terms $V_j(\vec r - \vec r_j)$, while having some finite spatial extent, are nevertheless radially symmetric about the point $r_j$, and that their Fourier transforms only weakly depend on the magnitude of the momenta $\vec k, \vec q$. This allows us to expand the equation above in angular channels. For this reason we introduce the quantities 
\begin{align}
\psi_{i}(\theta) &\equiv \int \frac{kdk}{2\pi} e^{i \vec k\cdot \vec r_i}\psi(\vec k) \nonumber \\
G_{ij}(E,\theta) &\equiv \int \frac{kdk}{2\pi} e^{i \vec k\cdot \left(\vec r_i - \vec r_j\right)}G_0(E,\vec k),  \nonumber
\end{align}
which yield the angular momentum components through the integrals $G_{ij}^l(E) = \int \frac{d\theta}{2\pi} G_{ij}(E,\theta) e^{-il\theta}$ and $\psi_i^l(E) = \int \frac{d\theta}{2\pi} \psi_i(E,\theta) e^{-il\theta}$. The spectral problem then takes the form 
\begin{equation}
\psi_i^l = \sum_j\sum_{l'} G_{ij}^{l-l'}(E)V_{j,l'}\psi_j^{l'},\label{eq:syseq}
\end{equation}
where the indices $l,l'$ label the angular momentum components and $i,j$ refer to the position indices of the nanomagnets. To solve the spectral problem \eqref{eq:syseq}, we must obtain expressions for the angular momentum components of the Green's function as well as the magnetic field. This derivation is done in the supplementary information (SI) \cite{si}, where the explicit forms of the results are also to be found. As shown there, the angular momentum components of the magnetic field are
\begin{equation}
V_{j,l} = 2\eta_j\sigma_z F_{j,l}/m,
\end{equation}
where $\eta_j = M_j\pi m R^2/\hbar^2$ is a coupling term and $F_{j,l}$ is given in terms of Bessel functions in the SI. As $|l|\to\infty$, for a fixed radius $R$ the terms $F_l$ vanish as $\propto l^{-(2l+1)}$, and hence above some $|l| > l_\mathrm{max}$ we can approximate $F_l = 0$. This effectively reduces the infinite number of equations to a finite one, and it is then straightforward to write Eq.~\eqref{eq:syseq} as the nonlinear matrix eigenvalue problem
\begin{equation}
\left[
 \beta^2
\begin{pmatrix}
A & 0\\
0 & 0
\end{pmatrix}
+ \beta
\begin{pmatrix}
V^{-1}  & B\\
B & V^{-1}
\end{pmatrix}
-
\begin{pmatrix}
0 & 0\\
0 & A
\end{pmatrix}\right] \Psi = 0,\label{eq:NLEVP}
\end{equation}
where $\beta = (\Delta + E)/\sqrt{\Delta^2-E^2}$. In the above, $A$ and $B$ are $4(2l_\mathrm{max}+1)\times 4(2l_\mathrm{max}+1)$ matrices with submatrix elements $A,B$ constructed from the Green's function, as detailed in the SI, and $V$ is a diagonal matrix constructed from $V_{j,l}$. One can regard Eq.~\eqref{eq:NLEVP} as a tight-binding problem for eigenvalues $E$ and eigenvectors $\Psi$. In contrast to ordinary tight-binding problems with a linear dependence on $E$, the energy dependence of the matrices in Eq.~\eqref{eq:NLEVP} is explicitly nonlinear.  To work around this problem, previous works have mainly focused on the mid-gap regime where the system can be linearized in $E$. This also allows one to derive an effective Hamiltonian and solve a normal linear eigenvalue problem \cite{pientka2,zhang}. Since the linear approximation would force us away from the physically most interesting parameter regime of large energy gaps and robust topological states, we will employ methods to treat the full nonlinear problem \cite{west,poyh2}. We note again that Eq.~\eqref{eq:NLEVP} is in principle valid for any configuration of radially symmetric magnets, and does not assume that they are placed in some particular lattice.

%
%
%
%
%
%
\begin{figure*}
\includegraphics[width=0.98\linewidth]{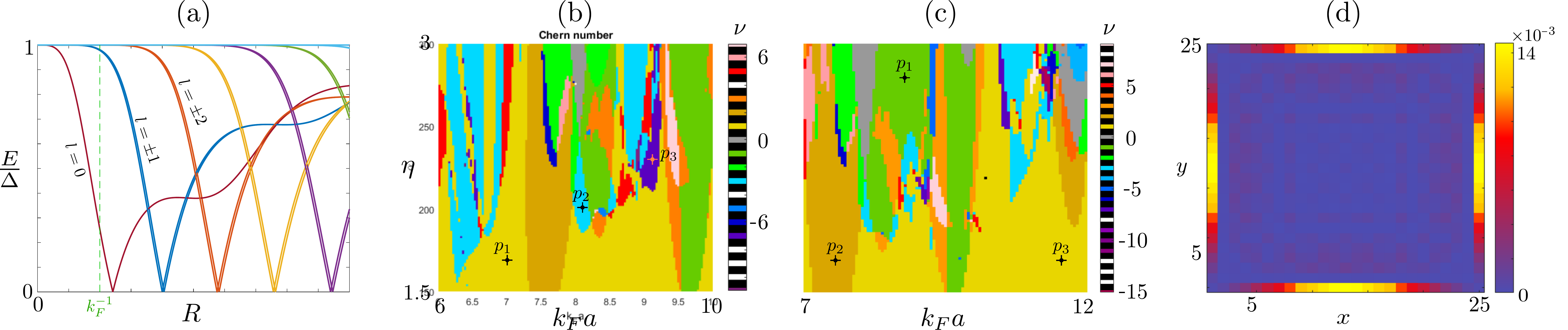}
\caption{(a) Bound state energies as a function of radius $R$ of a single magnet with constant magnetization. As $R \to 0$, the magnetic coupling vanishes, lifting the bound states to the gap. At $k_FR= 1$ all but the lowest three states are essentially gapped out. Parameters used are $\eta = 2$, $\varsigma = 0.1$  (b)  Chern number diagram calculated in real space for a $17\times 17$ system with $\xi/a = 2.5$, $R/a = 0.125$, $\varsigma=\alpha_R/v_F = 0.3$. The angular momentum cutoff is set at $l_\mathrm{max} = 1$. The energy gap at the selected points is $E(\vec p_1) \approx  0.15\Delta$, $E(\vec p_2) \approx  0.063\Delta$, $E(\vec p_3) \approx 0.060 \Delta$. (c) Similar diagram, but with $\xi/a = 2$, $R/a = 0.1$,  The energy gap at the selected points is $E(\vec p_1) \approx  0.11\Delta$, $E(\vec p_2) \approx 0.15\Delta$, $E(\vec p_3) \approx 0.27\Delta$. (d) Square of wavefunction amplitude, $|\psi|^2$, for the lowest-lying positive energy wavefunction for a $25\times 25$ lattice from the point $k_Fa = 7.5$, $\eta = 1.6$ in (c).}\label{Fig:all}
\end{figure*}

\textit{Single-magnet problem ---}
We first consider the case where a single magnet rests on the substrate. This reduces the matrix in the nonlinear eigenvalue problem to be diagonal in indices $i,j$. By taking the determinant of the matrix to be zero, we can find the solution for an \textit{arbitrary} value of $l_\mathrm{max}$ \cite{si}. In terms of $\beta$, the eigenvalues obtained are
\begin{widetext}
\begin{equation}
\beta_l =
\begin{cases}
\frac{\eta}{2}\left|F_{l} - F_{|l+1|}  \pm \sqrt{\left(F_{l} - F_{|l+1|}\right)^2 + \frac{4}{1 + \varsigma^2} F_{l}F_{|l + 1|}}\, \right|, &  0 \leq |l| < l_\mathrm{max}\\
\eta |F_{l}|, &|l| = l_\mathrm{max}
\end{cases}\label{eq:singlemag}
\end{equation}
\end{widetext}
where $\varsigma = \alpha_R/v_F$.
The bound-state energies are then obtained by using the relation $E_l = \Delta(\beta_l^2 -1)/(\beta_l^2 + 1)$. Essentially this constitutes a full solution of the bound state energies; the expressions for $|l| < l_\mathrm{max}$ do not depend on the cutoff, and so it can safely be taken to infinity. Note that as $\beta_{l}\underset{l \to \infty}{\longrightarrow} 0$, we have $E_{l}\underset{l \to \infty}{\longrightarrow} \Delta$, so beyond a certain $l$ the energies are effectively gapped out. 

For the purposes of engineering a gapped topological superconductor, a large number of gap-filling bound states presents a problem. The way of counteracting this would be making the nanomagnets sufficiently small that the lowest-energy states are well separated from the rest, if possible. The parameter controlling the number of relevant bound states is $k_FR$, where $k_F$ is the Fermi wavenumber of the 2DEG and $R$ is the radius of the magnet. Though small magnets pose a challenge to the fabrication process, in an experimental setting $k_FR \approx 1$ is already within reach since fabrication of nanomagnets with radius of a few tens of nanometers has become feasible \cite{menard}. With that in mind, in Fig.~\ref{Fig:all}(a) we plot the bound state energies as a function of $R$ for selected system parameters. As seen in the figure, while for larger $R$ the higher-$l$ states get increasingly important, for our parameters around $k_FR = 1$, only the lowest few states are appreciably within the gap, and the very lowest is separated from the others by a finite energy, which raises hopes for the presence of robust topological phases in realistic parameter regimes. Hence we expect that, for the studied parameter regime, a low value of $l_\mathrm{max}\lesssim 2$ should be an excellent approximation for the system with multiple magnets, since the low-lying states are unlikely to couple strongly to those near the gap edge.

The validity of this assumption can be readily tested by examining the properties of a one-dimensional chain of magnets, which we have done in the SI \cite{si}. We find that properties such as topology and energy gap at $l_\mathrm{max} = 1$ are essentially indistinguishable to those obtained for $l_\mathrm{max} = 2,\ldots,7$ in the studied regime. Based on this, we conclude that, for the values of $R$ and $k_F$ used here, $l_\mathrm{max} = 1$ is already a good approximation of the system.

\textit{ Two-dimensional lattices ---} Now we apply our theory to 2d systems. Guided by the single-magnet and 1d problems, we focus on the parameter regime where the angular momentum expansion can be cut at $l=1$. Even with this truncation the nonlinear eigenvalue problem in Eq.~\eqref{eq:NLEVP} involves $12N\times12N$ matrices, where $N$ is the number of magnets.  Working in $k$-space would reduce the dimension to a $12\times12$ problem, but the relevant Fourier transforms cannot be carried out analytically. This fact, and the large number of bands, makes analytical work intractable even in $k$-space. It is computationally more convenient to study the properties of finite systems in real space with periodic boundary conditions. In the considered parameter regime,  finite size properties converge rapidly even for relatively small systems.

To obtain the topological phase diagram we must also evaluate the Chern number for the system. Typically the calculation of the Chern number is formulated in momentum space, but it can be performed directly in real space. For this purpose, we will use the approach outlined in Ref.~\cite{zhang2}, requiring diagonalization of a real-space system with periodic boundary conditions, in a procedure briefly outlined in the SI \cite{si}. In Fig.~\ref{Fig:all}(b-c), we have plotted a topological phase diagram on a square lattice. As is seen in the figure, the selected parameter regimes support a wide range of topological phases, with Chern numbers varying from -15 to 9. Additional phases may be found by exploring other combinations of parameters. This abundance arises from the long-range inter-magnet coupling terms in the system, following the arguments in Ref.~\cite{ront1,ront2}. The spectral problem  can be solved numerically along the lines of Ref.~\cite{west,poyh}, though the large number of orbitals and the 2d nature of the system makes the present case computationally demanding. We have calculated the energy gap at a few selected points from Figs.~\ref{Fig:all} (b) and (c), listed in the figure caption. Notably, far from phase boundaries, systems in a nontrivial phase can have have energy gaps of the order of $0.25\Delta$ or higher, which could optimally translate to temperatures $T\sim 1$ K. In general, the energy gap decreases as the Chern number increases.  The Chern number of a 2D topological superconductor corresponds to the number of chiral edge modes around the system with open boundary conditions. The energies of the edge modes are located in the bulk excitation gap and provides an experimentally accessible fingerprint of the nontrivial topology. Indeed, as shown in \ref{Fig:all}(d), the states in the bulk gap of a nontrivial state are located on the edges of the sample.

It is important to address whether the parameter regime relevant to the system is feasibly achievable in experiment. As explained above, robust gapped states require that $k_FR\lesssim1$, where $R$ is the radius of the nanomagnet. Assuming that the radius of the nanomagnets is $R  = 50$ nm, it follows that, for example, the other parameters in Fig.~\ref{Fig:all}(c) are $\xi = 1$ $\mu$m, $k_F \approx 2\cdot 10^7$ m$^{-1}$; the characteristic energy scale of magnetization in Fig.~2 (b) and (c) is $M_i\sim 0.6 - 1.2$ meV. We compare this to two recent studies of InAs-based 2DEG-superconductor composite systems. References \cite{lee,hell} employ value $\Delta \approx 230 $ $\mu$eV. Using $m^* = 0.023 m_e$, $n_{2D} \approx 9\cdot 10^{15}$  m$^{-2}$ \cite{lee}, we obtain $k_F \approx 3.36\cdot 10^8$ m$^{-1}$, $\xi \approx 3.4$ $\mu $m. Furthermore, from the spin-orbit energy $\frac{m^*\alpha^2}{2} = 118.5$ $\mu$eV \cite{hell} we obtain $\varsigma \approx 0.15$. We conclude that our parameters are approximately in line with those studied, provided that the electron density $n_{2D}$ is reduced to $n_{2D} \approx 10^{14}$ m$^{-2}$.  In 2DEG materials the Fermi level can be gated even down to zero, which is complicated here by the screening from the proximity superconductor. However, while adding a technical difficulty similar to some previous proposals, for example Ref.~\cite{sau}, the presence of a superconductor does not pose a fundamental obstacle for electrostatic control of density. The superconductor does not need to be in a direct contact with the whole magnetic area to induce a robust proximity gap. Modern fabrication technology allows even quite imaginative solutions such as creating a checkerboard pattern with alternating magnetic and superconducting regions \cite{marcus}. The superconductivity may persist in proximity systems even for magnetic fields of several Teslas, so the system is expected to be robust against the local disruption due to the magnets. 

\textit{Discussion ---}There are two outstanding issues in the research of chiral topological superconductivity. The first one is the physical realization of chiral states in experimentally accessible systems, preferably in a way that allows a systematic study of states with distinct Chern numbers. The second one is to device a method that enables a unique identification of the Chern number of a state. Our work is a comprehensive effort toward the first goal. The second issue remains a challenge at the moment. The Majorana edge modes support a quantized thermal conductance determined by the number of modes which coincides with the Chern number. However, the required precision in the measurement of thermal transport is not feasible presently. While there exist proposals to identify the topological state through electric measurements \cite{ront1,rachel}, none of the known methods so far are general and practical enough to solve the problem satisfactorily. This is an area of active study and, due its versatility, the nanomagnet system is an excellent test bench for future proposals.    

Besides the rich topology, the crucial novelty of the studied system comes from hitherto unprecedented tuneability of the topological state. First of all, the structural control in the fabrication process enables controlling the lattice constant $a$ and the geometry of the magnetic array. Different stacking will modify the topological state and allow a fabrication of multiple different topological domains in one sample with chiral Majorana edge channels separating them. More importantly, the state of the system is tuneable by external control parameters after the fabrication. By tuning $k_F$ by electronic gates and magnetization through external fields, it is possible to sample the different regions of the phase diagram in Fig.~\ref{Fig:all}(b-c) \emph{in the same system}. A realization of the studied system requires state-of-the-art experimental efforts, which is natural for the proposed highly ambitious goal.

\textit{Conclusion ---}
In this work we have introduced a nanomagnet-semiconductor structure that serve as a universal platform for topological chiral superconductivity. This system supports several different topological states which can be  tuned by  structural design and electronic gates. The fabrication of the proposed system is within reach of current technology and could stimulate systematic research of mesoscopic superconductors with tuneable Chern numbers in the near future.

%
%
%
%
%
%

\textit{Acknowledgements ---\hspace{3mm}}
The authors would like to thank Charlie Marcus, Stevan Nadj-Perge and Pascal Simon for discussions. This work is supported by the Academy of Finland and the Aalto Centre for Quantum Engineering. K.P. acknowledges the Finnish Cultural Foundation for support.  

%
%
%
%
%
%

\begin{thebibliography}{16}

\bibitem{kallin} C. Kallin and J. Berlinsky, Rep. Prog. Phys. \textbf{79} 054502 (2016).
\bibitem{volovik} G. E. Volovik, \emph{The Universe in a Helium Droplet}, (Oxford University Press, 2003).
\bibitem{bernevig} B. A. Bernevig and T. L. Hughes, \emph{Topological Insulators and Superconductors}, (Princeton University Press, 2013).
\bibitem{schnyder} A. P. Schnyder, S. Ryu, A. Furusaki, and A. W. W. Ludwig, Phys. Rev. B \textbf{78}, 195125 (2008);  S. Ryu, A. P. Schnyder, A. Furusaki, and A. W. W. Ludwig, New J. Phys. \textbf{12}, 065010 (2010).
\bibitem{sau}J. D. Sau, R. M. Lutchyn, S. Tewari, and S. Das Sarma  Phys. Rev. Lett. \textbf{104}, 040502 (2010).
\bibitem{np2}S. Nadj-Perge, I. K. Drozdov, J. Li, H. Chen, S. Jeon, J. Seo, A. H. MacDonald, B. Andrei Bernevig, and Ali Yazdani, Science \textbf{346}, 602 (2014).
\bibitem{ruby} M. Ruby \emph{et al}., Phys. Rev. Lett. 115, 197204 (2015).
\bibitem{choy} T. P. Choy, J. M. Edge, A. R. Akhmerov, and C. W. J. Beenakker, Phys. Rev. B \textbf{84}, 195442 (2011).
\bibitem{np} S. Nadj-Perge, I. K. Drozdov, B. A. Bernevig, and A. Yazdani, Phys. Rev. B \textbf{88}, 020407(R) (2013).
\bibitem{brau1} B. Braunecker and P. Simon, Phys. Rev. Lett. \textbf{111}, 147202 (2013).
\bibitem{klin}J. Klinovaja, P. Stano, A. Yazdani, and D. Loss, Phys. Rev. Lett. \textbf{111}, 186805 (2013).
\bibitem{vazifeh} M.M. Vazifeh, M. Franz, Phys. Rev. Lett. \textbf{111}, 206802 (2013).
\bibitem{pientka2} F. Pientka, L. I. Glazman and F. von Oppen, Phys. Rev. B \textbf{88}, 155420 (2013).
\bibitem{poyh}K. P\"oyh\"onen, A. Weststr\"om, J. R\"ontynen, and T. Ojanen, Phys. Rev. B \textbf{89}, 115109 (2014).
\bibitem{heimes}  A. Heimes, P. Kotetes, G. Sch\"on, Phys. Rev. B \textbf{90}, 060507(R) (2014).
\bibitem{bry}  P. M. R. Brydon, S. Das Sarma, H.-Y. Hui, J. D. Sau, Phys. Rev. B \textbf{91}, 064505 (2015). 
\bibitem{ront1} J. R\"ontynen and T. Ojanen, Phys. Rev. Lett. \textbf{114}, 236803 (2015). 
\bibitem{ront2} J. R\"ontynen and T. Ojanen, Phys. Rev. B \textbf{93}, 094521 (2016). 
\bibitem{li1} J. Li, T. Neupert, Z. J. Wang, A. H. MacDonald, A. Yazdani, B. A. Bernevig, Nat. Comm. \textbf{7}, 12297 (2016).
\bibitem{yu} L. Yu, Acta Phys. Sin.  \textbf{21}, 75 (1965).
\bibitem{shiba} H. Shiba, Prog. Theor. Phys.  \textbf{40}, 435 (1968).
\bibitem{rusinov} A. I. Rusinov, JETP Lett.  \textbf{9}, 85 (1969).
\bibitem{balatsky}  A. V. Balatsky, I. Vekhter, and J.-X. Zhu,  Rev. Mod. Phys. 
\textbf{78}, 373 (2006).
\bibitem{kim} Y. Kim, J. Zhang, E. Rossi, and R. M. Lutchyn, Phys. Rev. Lett. \textbf{114}, 236804 (2015).
 \bibitem{zhang} J. Zhang, Y. Kim, E. Rossi, R. M. Lutchyn, Phys. Rev. B \textbf{93}, 024507 (2016).
\bibitem{west} A. Weststr\"om, K. P\"oyh\"onen, and T. Ojanen, Phys. Rev. B \textbf{91}, 064502 (2015).
\bibitem{poyh2}K. P\"oyh\"onen, A. Weststr\"om, and T. Ojanen, Phys. Rev. B \textbf{93}, 014517 (2016).
\bibitem{si}K. P\"oyh\"onen and T. Ojanen, Online Supplementary Information.
\bibitem{menard}G. C. M\'enard, S. Guissart, C. Brun, M. Trif, F. Debontridder, R. T. Leriche, D. Demaille, D. Roditchev, P. Simon, T. Cren, arXiv:1607.06353.
\bibitem{zhang2} Y. F. Zhang, Y. Y. Yang, Y. Ju, L. Sheng, D. N. Sheng, R. Shen, D. Y. Xing,  Chinese Phys. B \textbf{ 22}, 117312 (2013).
\bibitem{lee} J. S. Lee, B. Shojaei, M. Pendharkar, A. P. McFadden, Y. Kim, H. J. Suominen, M. Kjaergaard, F. Nichele, C. M. Marcus, C. J. Palmstrom, arXiv:1705.05049.
\bibitem{hell} M. Hell, M. Leijnse, and K. Flensberg, Phys. Rev. Lett. \textbf{118}, 107701 (2017).
\bibitem{marcus} C. Marcus, in private communication.
\bibitem{rachel}S. Rachel, E. Mascot, S. Cocklin, M. Vojta, D. K. Morr, arXiv:1705.05378. 

\end{thebibliography}

%
%
%
%
%
%
%
%
%
%
%
%
%
%
%
%
%
%
%
%
\newpage
\begin{widetext}
\begin{center}
\Large SUPPLEMENTAL INFORMATION
\end{center}
\appendix
\section{Green's function integrals}
In this section, we explicitly calculate the terms in the angular momentum expansion of the system Green's function. These are defined by the equation
\begin{equation}
G_{ij}^l(E) = \int \frac{d\theta}{2\pi} G_{ij}(E,\theta) e^{-il\theta}.\label{App:eq:glijdef}
\end{equation}
To begin with, we need an explicit expression for the function in the integrand, which can be obtained through 
\begin{equation}
G_{ij}(E,\theta) \equiv \int \frac{kdk}{2\pi} e^{i \vec k\cdot \left(\vec r_i - \vec r_j\right)}G_0(E,\vec k).
\end{equation}
By using projectors to the eigenstates of the SOC term, the Green's function can be written as a sum over two helicities:
\begin{equation}
\begin{split}
G_{ij}(E,\theta) &= \frac{1}{2}\sum_{\lambda = \pm 1} \left[1 + \lambda(-i\sigma_-e^{i\theta} + i\sigma_+e^{-i\theta})\right]\int\frac{kdk}{2\pi} e^{i k r_{ij}\cos(\theta - \theta_{ij})}
\frac{E + \xi_\lambda\tau_z + \Delta\tau_x}{E^2 - \xi_\lambda^2 - \Delta^2}
\end{split}
\end{equation}
where $\sigma_\pm = \frac{1}{2}(\sigma_x\pm i\sigma_y)$, $\xi_\lambda=v_F(k-k_F^\lambda)$, $k_F^\lambda=k_F(1+\lambda\varsigma\sqrt{1+\varsigma^2})$, and we have introduced the normalized spin-orbit coupling $\varsigma \equiv \alpha_R/v_F$  and the shorthand $\vec r_{ij} = \vec r_i - \vec r_j$; we also let $\theta_{ij}$ denote the angle $\vec r_{ij}$ makes with the $x$ axis. We obtain
\begin{align}
G_{ij}(E,\theta)
\approx \frac{m}{2}&\sum_{\lambda = \pm 1} e^{i k_F^\lambda r_{ij}\cos(\theta - \theta_{ij})}\left[1 - \tfrac{\lambda\varsigma}{\gamma}\right]\left[1 + i\lambda(-\sigma_-e^{i\theta} + \sigma_+e^{-i\theta})\right]\left[(E+ \Delta\tau_x)g_1 + \tau_zg_2\right],
\end{align}
where we have adopted the shorthand $\gamma \equiv \sqrt{1 + \varsigma^2}$. The integrals above are
\begin{align}
g_1 &= \int_{-\infty}^\infty\frac{d\xi}{2\pi} e^{i \xi r_{ij}\cos(\theta - \theta_{ij})/(\gamma v_F)}
\frac{1}{E^2 - \xi_\lambda^2 - \Delta^2}\\
g_2 &= \int_{-\infty}^\infty\frac{d\xi}{2\pi} e^{i \xi r_{ij}\cos(\theta - \theta_{ij})/(\gamma v_F)}
\frac{\xi_\lambda}{E^2 - \xi_\lambda^2 - \Delta^2}
\end{align}
Both integrals can be solved through standard residue integration,
yielding
\begin{align}
g_1 &= \frac{e^{- \frac{r_{ij}}{\xi_E}|\cos(\theta - \theta_{ij})|}}{2\sqrt{\Delta^2 - E^2}}\\
g_2 &=
\begin{cases}
0, & r_{ij} = 0\\
-\frac{i}{2}\sgn(\cos(\theta - \theta_{ij}))
e^{- \frac{r_{ij}}{\xi_E}|\cos(\theta - \theta_{ij})|}, &
r_{ij} \neq 0
\end{cases}
\end{align} 
where the energy-dependent coherence length has been defined as 
$\xi_E \equiv \gamma v_F/\sqrt{\Delta^2 - E^2}$.

Having obtained $G_{ij}(E,\theta)$, the next step is to find the angular momentum components of the same, as defined by Eq.~\eqref{App:eq:glijdef}. Upon inspection this results in the equation
\begin{equation}
G^{l}_{ij}(E) =
-\frac{m}{4}\sum_{\lambda = \pm 1} \times
\begin{cases}
\frac{E + \Delta\tau_x}{\sqrt{\Delta^2 - E^2}}\left[\delta_{l,0} + i \tfrac{\varsigma}{\gamma}(\sigma_-\delta_{l,1} - \sigma_+\delta_{l,-1})\right], &\vec r_i = \vec r_j\\
\bigg(\frac{E + \Delta\tau_x}{\sqrt{\Delta^2 - E^2}}\left[A^l_\lambda(\vec r_{ij}) - i\lambda\sigma_- A^{l-1}_\lambda(\vec r_{ij}) + i\lambda\sigma_+A^{l+1}_\lambda(\vec r_{ij}) \right]\notag\\
\hspace{2.8cm}+\left[B^l_\lambda(\vec r_{ij}) - i\lambda\sigma_- B^{l-1}_\lambda(\vec r_{ij})  + i \lambda\sigma_+B^{l+1}_\lambda(\vec r_{ij}) \right]
 \bigg), &\vec r_i\neq \vec r_j
\end{cases}
\end{equation}
where the case $r_{ij} = 0$ is immediately seen, and the integrals in the $r_{ij} \neq 0$ case are defined by
\begin{align}
A^l_\lambda(\vec r_{ij}) &=
\left(1 - \tfrac{\lambda\varsigma}{\gamma}\right)e^{-il\theta_{ij}}\int_0^{2\pi} \frac{d\theta}{2\pi}e^{-il\theta} e^{i k_F^\lambda r_{ij}\cos\theta - \frac{r_{ij}}{\xi_E}|\cos\theta|}\\
B^l_\lambda(\vec r_{ij}) &=
i\left(1 - \tfrac{\lambda\varsigma}{\gamma}\right)e^{-il\theta_{ij}}\int_0^{2\pi} \frac{d\theta}{2\pi}\sgn(\cos\theta)e^{-il\theta} e^{i k_F^\lambda r_{ij}\cos\theta - \frac{r_{ij}}{\xi_E}|\cos\theta|}.
\end{align}
These can more conveniently be written in the form
\begin{align}
A^l_\lambda(\vec r_{ij}) &= 2\left(1 - \tfrac{\lambda\varsigma}{\gamma}\right)\left[\frac{y_{ij} + ix_{ij}}{r_{ij}}\right]^l\RE\left[(-i)^lI_l(z^\lambda_{ij})\right]\\
B^l_\lambda(\vec r_{ij}) &= -2\left(1 - \tfrac{\lambda\varsigma}{\gamma}\right)\left[\frac{y_{ij} + ix_{ij}}{r_{ij}}\right]^l\IM\left[(-i)^lI_l(z^\lambda_{ij})\right],
\end{align}
where $z^\lambda_{ij} \equiv (k_F^\lambda + \frac{i}{\xi_E})r_{ij}$ and the remaining integral is
\begin{equation}
I_l(z) \equiv \int_{-\frac{\pi}{2}}^{\frac{\pi}{2}} \frac{d\theta}{2\pi}e^{-il\theta + i z\cos\theta}.
\end{equation}
The value of the integral for a given $l$ can be obtained through recurrence:
\begin{align}
I_l(z) &= \frac{2i(l-1)}{z}I_{l-1}(z) + I_{l-2}(z) - \frac{i^{l-1}}{z}\left[(-1)^l + 1\right]\\
I_0(z) &= \frac{1}{2}\left[J_0(z) + iH_0(z)\right]\\
I_1(z) &= \frac{1}{\pi} + \frac{i}{2}\left[J_1(z) + iH_1(z)\right].
\end{align}
Using the above $2\times 2$ matrices $A^l_\lambda(\vec r_{ij}), B^l_\lambda(\vec r_{ij})$ it is then straightforward to construct the $2(2l_\mathrm{max} + 1)\times 2(2l_\mathrm{max} + 1)$ matrices $A,B$ in the main text by making the angular momentum sum into a matrix equation; the specific form in Eq.~\eqref{eq:NLEVP} is obtained through a unitary transformation to the $\tau_x$ basis.

\section{Components of magnetic potential}
The goal in this section is to obtain the angular momentum components of the magnetic potential, assuming the potential in real space is described by the equation
\begin{equation}
V_j(\vec r) = M_j\sigma_j\theta(R_j -|\vec r|).
\end{equation}
We calculate the Fourier transform $V_j(\vec k - \vec q)$ in the vicinity of the Fermi surface,  so we can assume that the magnitudes of the wave vectors are equal to the Fermi wavevector $k_F$:
\begin{equation}
\begin{split}
V(\vec k,\vec q) &\approx M_j\sigma_j\int_0^{2\pi}\! d\theta \int_0^{R_j}  rdr\\
&\qquad\quad \times
\left[e^{-ik_Fr[\cos(\theta_k-\theta) - \cos(\theta_q-\theta)]}\right].
\end{split}
\end{equation}
By expanding the exponent in Bessel functions and performing the integral over $\theta$, we obtain
\begin{equation}
V_j(\theta_{kq}) = 2\pi M_j\sigma_j \sum_{l = -\infty}^\infty e^{il\theta_{kq}} \int_0^{R_j} drJ_l(k_F r)^2 r ,
\end{equation}
and, hence,
\begin{equation}
V_{j,l} = \frac{2}{m}\eta_j\sigma_j F_{j,l},
\end{equation}
where $\eta_j = M_j\pi m$ and
\begin{align}
F_{j,l} &= \int_0^{R_j} drJ_l(k_F r)^2 r
\end{align}
This can be solved to obtain
\begin{align}
F_{j,l} &= \frac{1}{2}R_j^2\bigg[J_l(k_F R_j)^2 - \frac{2l}{k_FR_j}J_l(k_F R_j)J_{l+1}(k_F R_j) + J_{l+1}(k_F R_j)^2\bigg]
\end{align}

\section{From NLEVP to solutions}
The central equation of our system and the starting point for further analysis is Eq.~\eqref{eq:NLEVP} in the main text,
\begin{equation}
\left[\beta^2 M_2(E) + \beta M_1(E) + M_0(E)\right]\Psi = 0,
\end{equation}
where $\beta = (\Delta + E)/\sqrt{\Delta^2-E^2}$ and $\Psi$ is a vector of components
\begin{equation}
\psi^l_i = \int \frac{d\theta}{2\pi} \psi_i(E,\theta) e^{-il\theta}.
\end{equation}
This constitutes a $4N(2l_\mathrm{max}+1)\times 4N(2l_\mathrm{max}+1)$ nonlinear eigenvalue problem for the energy. We will utilize two approaches in solving this problem: first, we notice that for the typical subgap energy scales of the problem the matrices $M_i$ depend only weakly on $E$ through the energy-dependent coherence length $\xi_E$. Hence, as investigated in the Refs. [26-27] in the main text, to good accuracy it is sufficient to solve the \textit{polynomial} eigenvalue problem
\begin{equation}
\left[\beta^2 M_2(0) + \beta M_1(0) + M_0(0)\right]\Psi = 0\label{app:eq:heff}
\end{equation}
instead. The accuracy of this approach has been proven excellent and is employed in solving for the energy and wavefunctions. For the purpose of extracting the topological properties of the system, we define the topological Hamiltonian
\begin{equation}
\tilde H = M_2(0) +  M_1(0) + M_0(0).
\end{equation}
This will not in general have the same eigenvalues or eigenvectors as the full problem, and is hence not useful for the purposes of obtaining those. However, the topological properties of the system described by this Hamiltonian are equivalent to those described by the NLEVP. Notably, the topological Hamiltonian yields exact zero energy wavefunctions of the full problem and provides asymptotically accurate approximations in the vicinity of $E=0$ where topological phase transitions take place. This is sufficient for calculating the topological phase diagram everywhere since one only requires information near the transitions.  Hence $\tilde{H}$ can be conveniently used for obtaining topological invariants.

\section{The single-magnet problem}
In a system consisting of a single magnet, $G^l_{ij}$ vanishes for $|i-j| > 1$, and Eq.~\eqref{eq:syseq} from the main text can be written out explicitly as
\begin{equation}
\psi^l = -\eta\frac{E + \Delta\tau_x}{\sqrt{\Delta^2 - E^2}}\left[F_l\sigma_z\psi^l + i \tfrac{\varsigma}{\gamma}\sigma_-F_{l-1}\psi^{l-1} + i \tfrac{\varsigma}{\gamma}\sigma_+F_{l+1}\psi^{l+1}\right].
\end{equation}
The above can be immediately diagonalized in $\tau$ space by moving to the $\tau_x$ eigenbasis. Doing so (while still denoting the wavefunction $\psi$) results in the equation
\begin{equation}
\psi^l = -\eta\frac{E + \Delta\tau_z}{\sqrt{\Delta^2 - E^2}}\left[F_l\sigma_z\psi^l + i \tfrac{\varsigma}{\gamma}\sigma_-F_{l-1}\psi^{l-1} + i \tfrac{\varsigma}{\gamma}\sigma_+F_{l+1}\psi^{l+1}\right].
\end{equation}
At this point we can introduce an angular momentum cutoff $l_\mathrm{max}$, and define the vector
\begin{align*}
\Psi = (\psi^{-l_\mathrm{max}}, \ldots, \psi^0, \ldots, \psi^{l_\mathrm{max}})
\end{align*}
This allows the above to be written as a matrix equation for $E$ and $\Psi$. If we introduce the matrices $D$, $U$, and $L$, defined by
\begin{align}
D_{mn} &= F_{|l_{max} - m + 1|}\delta_{mn}\\
U_{mn} &= \frac{i\varsigma}{\gamma}F_{|l_{max} - n + 1|}\delta_{m+1, n}\\
L_{mn} &= \frac{i\varsigma}{\gamma}F_{|l_{max} - n + 1|}\delta_{m-1,n},
\end{align}
where $m,n = 1,\ldots, 2l_\mathrm{max}+1$, we have
\begin{align}
\begin{pmatrix}
1 + \eta \beta\left[\sigma_zD + i\tfrac{\varsigma}{\gamma}(\sigma_- U + \sigma_+ L)\right] & 0\\
0 & 1 + \eta \beta^{-1}\left[\sigma_zD + i\tfrac{\varsigma}{\gamma}(\sigma_- U + \sigma_+ L)\right]
\end{pmatrix}\Psi = 0.
\end{align}
In the above, $\beta \equiv (\Delta + E)/\sqrt{\Delta^2-E^2}$, and energy dependence is solely through $\beta$. This is a polynomial eigenvalue problem and can be solved by requiring that the determinant of the matrix be zero. Noting that $\beta$ and $\beta^{-1}$ yield the same energy $\pm E$ up to sign, to find all energies it is sufficient to solve
\begin{equation}
\det\left(1 + \eta \beta^{-1}\left[\sigma_zD + i\tfrac{\varsigma}{\gamma}(\sigma_- U + \sigma_+ L)\right]\right) = 0
\end{equation}
We multiply both sides by $\beta$ and write out the matrix structure in spin space explicitly:
\begin{equation}
\det
\begin{pmatrix}
\beta + \eta D & L\\
\eta U & \beta - \eta D
\end{pmatrix}
= 0
\end{equation}
The matrices $\beta \pm \eta D$ are diagonal. As $\beta > 0$, $F_l > 0$ we conclude that $\beta + \eta D$ must be nonsingular and write
\begin{equation}
\det\left[\beta + \eta D \right]\det\left[\beta - D - \eta^2 U (\beta + \alpha D)^{-1} L \right] = 0.
\end{equation}
The left-hand determinant is nonvanishing and can be divided away. We denote temporarily $(\beta + \alpha D)^{-1} \equiv M$, which is a diagonal matrix. Now
\begin{align}
\left[UML\right]_{ij} &= \sum_{mn} U_{im}M_{mn}L_{nj}\delta_{mn}\notag\\
&= -\frac{\varsigma^2}{\gamma^2}\sum_{n} F_{|l_{max} - n + 1|}\delta_{i+1,n}M_{nn}F_{|l_{max} - j + 1|}\delta_{n-1,j}\notag\\
&= -\frac{\varsigma^2}{\gamma^2}F_{|l_{max} - i|}M_{i+1,i+1}F_{|l_{max} - j + 1|}\delta_{j+1,i+1}
\end{align}
which is diagonal, with a zero on the diagonal at $i=j = 2l_{max} + 1$. Hence, taking the determinant of the matrix results in $2l_{max} + 1$ equations, one for each element on the diagonal. First, if $j < 2l_{max} + 1$, we have
\begin{align}
\beta - \eta F_{|l_{max} - j + 1|}  &=  -\frac{\eta^2 \varsigma^2}{\gamma^2}\frac{F_{|l_{max} - j|}F_{|l_{max} - j + 1|}}{\beta + \eta F_{|l_{max} - j|}}\notag
\end{align}
which can be solved to get
\begin{align}
\beta &=  \frac{\eta}{2}\left[F_{|l_{max} - j + 1|} - F_{|l_{max} - j|}\right] + \frac{\eta}{2}\sqrt{\left[F_{|l_{max} - j + 1|} - F_{|l_{max} - j|}\right]^2 + \frac{4}{\gamma^2} F_{|l_{max} - j|}F_{|l_{max} - j + 1|}}.
\end{align}
In addition to the $2l_\mathrm{max}$ solutions above, the $(2l_\mathrm{max}+1)$th equation yields upon inspection
\begin{equation}
\beta = \alpha F_{l_{max}}.
\end{equation}
Relabeling of indices finally yields Eq.~\eqref{eq:singlemag} in the text, when taking into account that $F_{-l} = F_l$.

\section{The one-dimensional chain and the $l$ cutoff}

In this section of the supplemental material, we study the properties of a one-dimensional chain of circular, homogeneous magnets on a 2DEG-superconductor substrate. This system is described by Eq.~\eqref{eq:syseq}, with magnet positions $\vec r_j = (x_j,0)$. 

As discussed in the main text, we introduce an angular momentum cutoff to obtain the more tractable Eq.~\eqref{eq:NLEVP} rather than attempting a full solution of Eq.~\eqref{eq:syseq}. For a chain of length $L$ this reduces the problem to a $4L(2l_\mathrm{max} + 1)\times 4L(2l_\mathrm{max} + 1)$ nonlinear eigenvalue problem. In the main text, we argued that a low value for the cutoff $l_\mathrm{max}$ should suffice as good approximation, based on the rapid convergence of $F_l$ as $l\to \infty$ as well as the results from the single-magnet problem. We can test this explicitly by solving for the energy eigenvalues of a finite one-dimensional chain at different cutoffs and comparing the results. The result of this is seen in Fig.~\ref{fig:app:lmax}, where we show a plot of the sorted eigenvalues for different values of $l_\mathrm{max}$. As seen in the figure, which is representative of our parameter regime, the eigenvalues obtained at $l_\mathrm{max} = 1$ are practically indistinguishable from those obtained with higher cutoffs; the relative difference between $l_\mathrm{max} = 1$ and $l_\mathrm{max} = 7$ is less than one percent. In general, a larger product $k_FR$ will necessitate the inclusion of more angular momentum states to get a quantitatively reliable approximation.

Having established that $l_\mathrm{max} = 1$ is a good approximation within our parameter regime, we can move on to study the topology of the magnet chain.
The chain belongs to the symmetry class BDI with chiral symmetry. This implies that there exists a unitary matrix $\mc C$ which anticommutes with the topological Hamiltonian $\tilde{H}$ Eq.~\eqref{app:eq:heff}. The topological states of a chain is given by the winding number invariant
\begin{equation}
\nu = \frac{1}{4\pi i} \int_{-\frac{\pi}{a}}^{\frac{\pi}{a}} dk \TR\left[\mc C \tilde{H}^{-1} \partial_k \tilde{H}\right].
\end{equation}
In general, consistent with the analogous Yu-Shiba-Rusinov system [27], we find a total of 5 phase with $|\nu|\leq 2$; see Fig.~\ref{Fig:App:Winding}(a), where we have plotted the winding number as a function of $k_F$, $\eta$ for a chain using $l_\mathrm{max} = 1$. As seen in the figure, there are some parameter regimes in which the winding number is not properly quantized; this is due to the very low energy gap in this regions, which makes a numerical calculation unreliable and the actual phase too susceptible to perturbations to be relevant for experimental realizations. This is seen in Fig.~\ref{Fig:App:Winding}(b) where we plot an energy gap diagram with the same parameters.

\begin{figure}
\includegraphics[width = 0.47\linewidth]{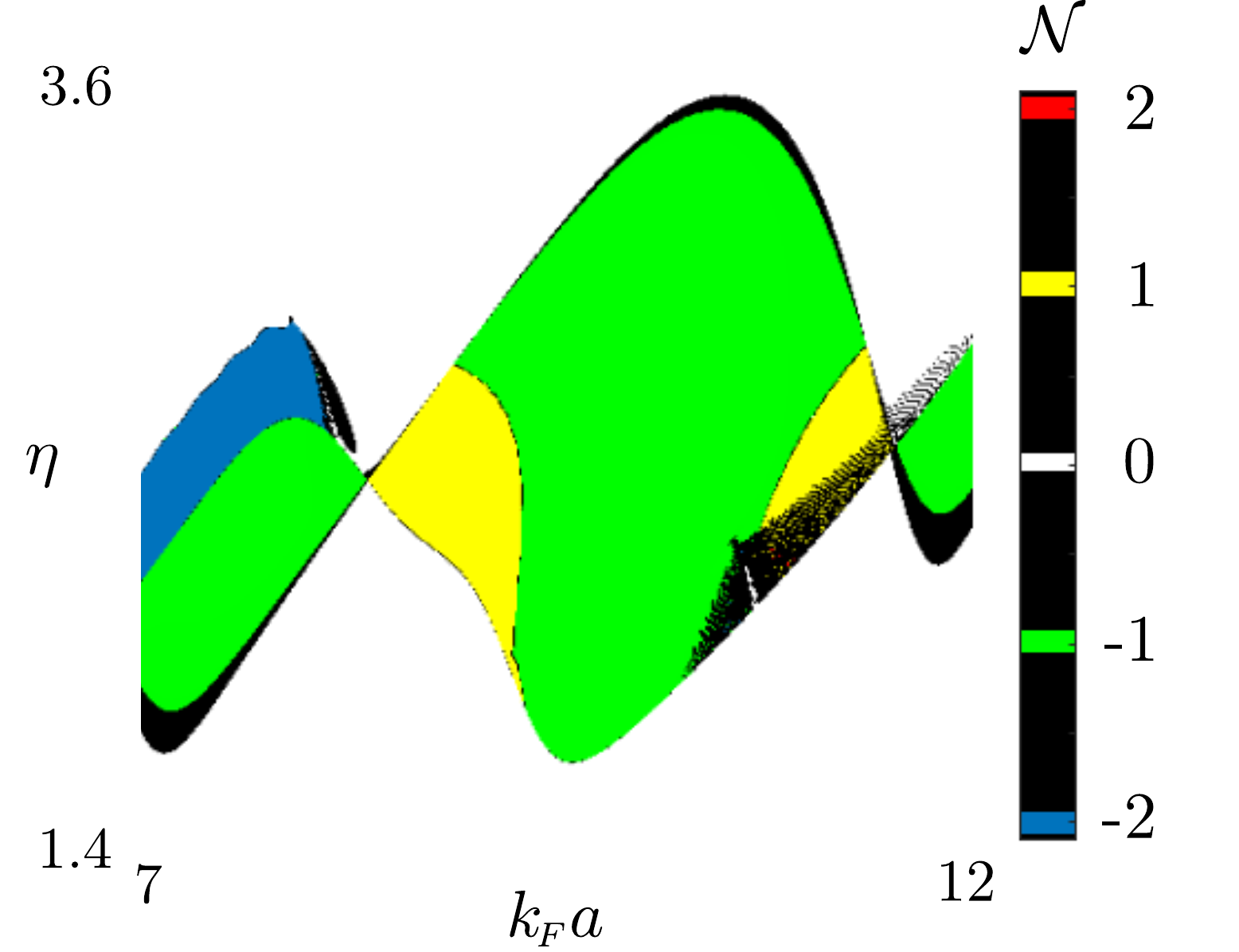}
\includegraphics[width = 0.47\linewidth]{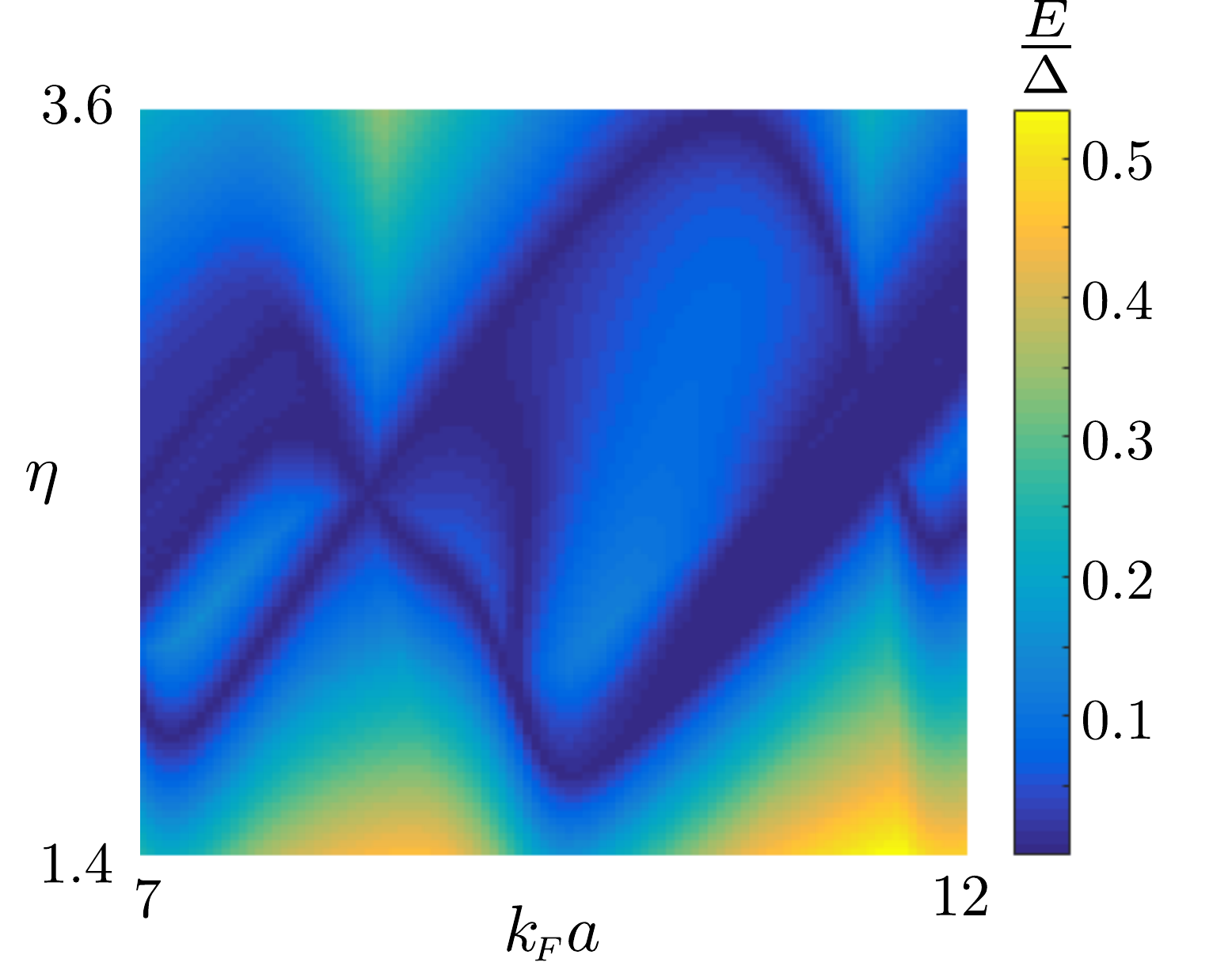}
\caption{Topological phase diagram of the one-dimensional nanomagnet chain. (a) Winding number diagram at $l_\mathrm{max} = 2$, with parameters used $\varsigma = 0.3$, $R = 0.1a$, $\xi = 2a$. (b) Energy gap diagram for the same parameters.}
\label{Fig:App:Winding}
\end{figure}

\begin{figure}
\includegraphics[width=\linewidth]{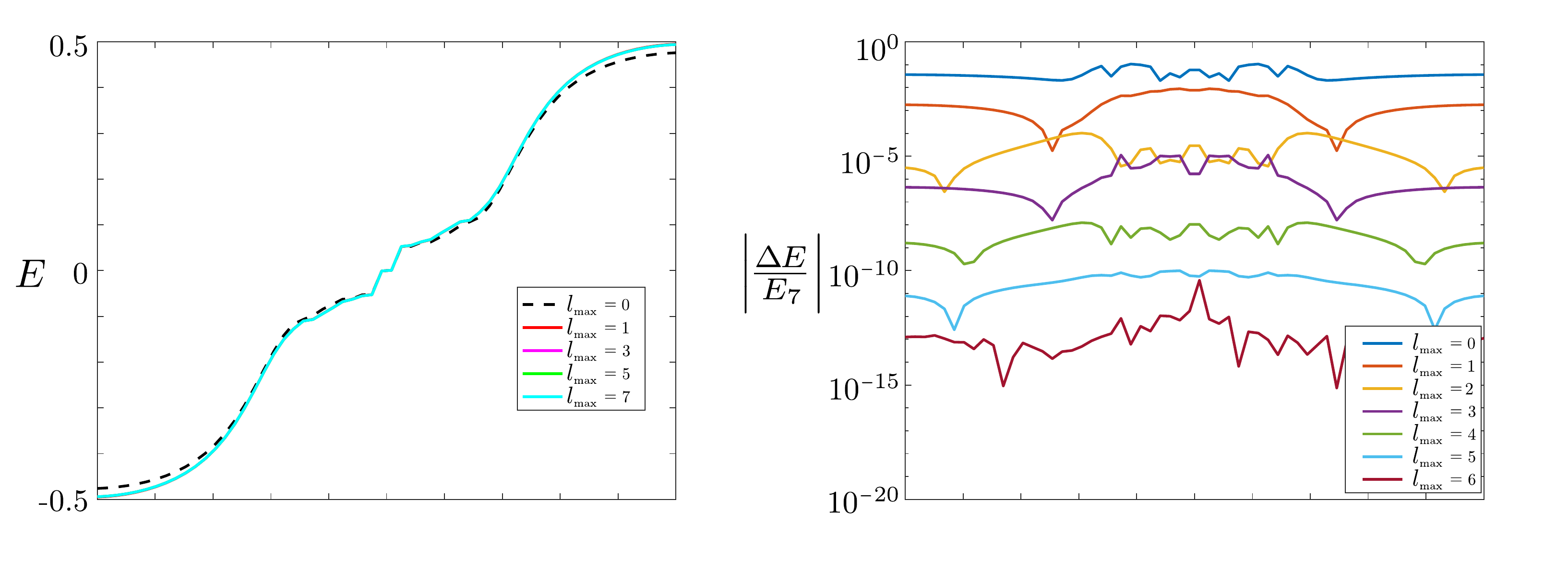}
\caption{(a)Sorted energy eigenvalues for a one-dimensional chain of magnets for $l_\mathrm{max} = 0,1,3,5,7$. As is seen in the figure, the $l_\mathrm{max} = 1,\ldots,7$ eigenvalues are largely overlapping, while the $l_\mathrm{max} = 0$ values do differ slightly. For the purposes of this figure, eigenvalues at the superconducting gap have been excluded. (b) Absolute relative deviation from the $l_\mathrm{max} = 7$ energy, $|E_{l_\mathrm{max}} - E_7|/|E_7|$, for $l_\mathrm{max} = 1,\ldots,6$. }\label{fig:app:lmax}
\end{figure}

\section{Real-space Chern number}
A standard way to obtain the Chern number is by obtaining the band projectors of the occupied states in $k$-space and using the relation
\begin{equation}
\mc C = \frac{i\epsilon_{0\mu\nu}}{2\pi}\int d^2p \TR\left[P_- \partial_{p_\mu}P_-\partial_{p_\nu}P_-\right].
\end{equation}
This is, however, more convenient for systems with a simpler matrix structure in which the projectors for each $k$-point can be solved analytically. Depending on the chosen cutoff $l_\mathrm{max}$, however, the $k$-space effective Hamiltonian considered here can easily be too large for that to be a feasible approach, which necessitate a numerical solution for each pair $(k_x,k_y)$. This can be bypassed by instead calculating the Chern number in real space, which is more computationally efficient for low values of $\xi/a\sim1-10$ (where smaller systems are viable). One method is presented in Ref.~[29]: define the coupling matrices $C_{\alpha,\alpha+1}$, with elements
\begin{equation}
C^{mn}_{\alpha,\alpha+1} = \bra{\psi^m}e^{i(\vec q_\alpha - \vec q_{\alpha + 1})\cdot \vec r} \ket{\psi^n},
\end{equation}
where $\vec q_\alpha = \pi (\delta_{\alpha,1} + \delta_{\alpha,2},\delta_{\alpha,2} + \delta_{\alpha,3})$, and where $\psi^m$ are the eigenfunctions of the system with periodic boundary conditions. By use of these matrices, the Chern number is then obtained through the equation
\begin{equation}
\mc C = \frac{1}{2\pi} \sum \arg(\lambda_m),
\end{equation}
with $\lambda_m$ being the complex eigenvalues of the matrix $C_{01}C_{12}C_{23}C_{30}$.
\end{widetext}
\end{document}